\documentclass{mem}
\usepackage{natbib}\usepackage{txfonts}\usepackage{balance}
\usepackage{graphicx}
\usepackage[a4paper,breaklinks,dvipdfm]{hyperref}
\idline{75}{282}
\begin{document}
\def\teff{$T\rm_{eff }$}
\newcommand{\kms}{\ensuremath{\mathrm {km\,s}^{-1}}}

\title{Runaway and hypervelocity stars}
\subtitle{The supernova connection}

\author{
R. Napiwotzki\inst{1} 
\and M. D. V. Silva\inst{1}
          }

  \offprints{R. Napiwotzki}

\institute{
Centre for Astrophysics Research, STRI, University of Hertfordshire, 
College Lane, Hatfield AL10 9AB, UK
\email{r.napiwotzki@gmail.com}
}

\authorrunning{Napiwotzki \& Silva}

\titlerunning{Runaway and hypervelocity stars}

\abstract{We present an investigation of the known sample of runaway
  stars. The orbits of these stars are traced back to their origin in the
  Galactic disc. The velocity distribution of these stars is compared to
  theoretical predictions. We conclude that the majority of stars is well
  explained by the standard binary ejection mechanism (BEM) and the dynamical
  ejection mechanism (DEM). However, we find a sample of ten stars which has
  ejection velocities in excess of those predicted by standard scenarios. We
  discuss how these can be explained by a variant of the BEM. This mechanism
  can create runaway stars exceeding the Galactic escape velocity (known as
  hypervelocity stars). The number of runaway stars in our Galaxy is estimated
  and compared to the known sample of high mass X-ray binaries, whose
  formation is linked to the BEM channel.
   } 
\maketitle{}

\section{Introduction}

In the commonly accepted picture, star formation in our Galaxy is confined to
the star forming regions in the Galactic disc. Runaway stars are young,
early-type stars observed outside young OB associations and open clusters.
Starting with the seminal work of \citet{G.S1974} the existence of early-type
main sequence (MS) stars in the Galactic halo is now well established. Star
formation in situ in the halo is an intriguing possibility, but ejection of
young stars from the disc is a possible alternative. The two channels under
discussion are the  binary
ejection mechanism (BEM) and the dynamical ejection mechanism (DEM).

In the BEM scenario \citep{Bla1961} the
(initially), higher mass star of the binary explodes as a supernova (SN). The
lion's share of the mass of the SN progenitor is ejected at high speed and
leaves the system within a short interval of time -- short compared to the
orbital period. Now, the remaining MS star feels a much diminished gravitation
pull from the SN remnant (usually a neutron star, NS). 
Depending on the mass ratio
and the orbital parameters of the progenitor system this can already be enough
to break up the binary. However, it has become clear that NS born
in a SN explosion receive an extra kick, which can amount to several
100\,\kms\ -- reducing the chance of a binary surviving the SN explosion even
further.  In the BEM scenario 
the surviving MS star should leave its place of birth
with a ejection velocity amounting to approximately its original orbital
velocity, which in the standard scenario can reach values $\la 300$\,\kms\
\citep{L.D1993, PS2000}.

An alternative is the DEM scenario proposed by
\citet{P.R.A1967}. Close encounters (``collisions'') between stars in young, 
dense clusters can result in one or both of them being ejected from the
cluster with collisions between two
binaries being the most efficient mechanism to produce large ejection
velocities. The DEM predicts ejection velocities up to
300--400\,\kms\ \citep{Leo1991, G.G.PS2009}. 

High mass X-ray binaries (HMXB) consists of a massive OB star with an
NS (or black hole) companion. They can be interpreted as those
systems which survived the SN explosion in the BEM scenario. \citet{C.I1998}
determined tangential velocities of a sample of HMXB finding a relatively high
sample average for systems with an OB supergiant component
$\overline{v_{\mathrm{tan}}} = 42\,\kms$ while the value for systems with Be
components was $\overline{v_{\mathrm{tan}}} = 15\,\kms$ \citep[corrected values
from ][]{vdH.PS.B2000}. 

This is consistent with expectations from the BEM. It is easier for systems
with higher mass secondaries (the OB supergiant progenitors) to survive a
modest kick. Note that this kick results largely from the specific momentum of
the SN ejecta leaving the system with the NS kick being only a minor
contribution.  It is interesting to note that a number of HMXBs fulfils the 
definition
of runaway stars -- a peculiar velocity in excess of 30\,\kms relative to
their standard of rest (SoR). However, very large kick
velocities would be incompatible with the survival of the system. 

The most basic test of whether MS runaway stars in the halo are explained by
the BEM and DEM scenarios is to reconstruct their
trajectory and check whether their flight times are compatible with a disc 
origin (i.e.\ not longer than their evolutionary lifetime). A further test is
the comparison of observed velocity distribution with predicted once. We will
describe an investigation of the known sample of runaway stars in
Sect.~2. Links with hypervelocity stars and HMXBs will be discussed in 
Sects.~3 and~4.

\section{The runaway stars}

Investigations of runaway stars face the obstacle that they are easily
confused with hot evolved stars (mainly sub\-dwarf~B stars and post-AGB
stars). Good quality spectra are necessary to distinguish the different types
of objects. We surveyed the literature on known runaway candidates at high
Galactic latitude and created an initial sample of 174 stars for which data of
sufficient quality were available \citep{S.N2011}. We used criteria as the
position in the Hertzsprung--Russell diagram, observed abundances and rotation
velocities to evaluate their status. This process left a sample
of 96 bona fide runaway stars for further analysis.

The location of the remaining stars in the Galactic $UVW$ coordinate system
was determined using their position in the sky and their distance
computed using $T_{\mathrm{eff}}$ and $\log g$ from spectroscopic and
photometric analyses and interpolating their mass in theoretical main sequence
tracks. Knowing the distance the 3D space velocity can be evaluated from the
measured radial velocity and proper motions. These are all ingredients needed
to reconstruct the orbit of the runaways stars back to their point of origin
in the Galactic disc using an updated version of the Galactic potential of
\citet{A.S1991}. We carried out a rigorous error analysis using a Monte Carlo
procedure \citep[see][for details]{S.N2011}.  

\begin{figure*}[t!]
{\centering
\hspace{0.1\hsize}\resizebox{0.8\hsize}{!}{\includegraphics[clip=true]{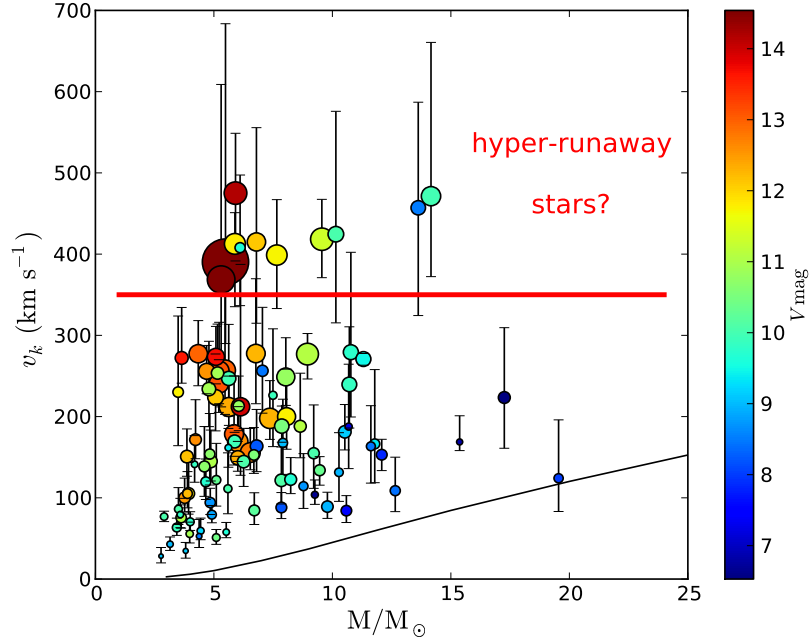}}
}
\caption{\footnotesize
Ejection velocities of the runaway stars as function of mass. The colour scale
indicates the observed brightness. The size of the circles is proportional to
the height above the Galactic plane. The bottom line indicates the minimum 
velocity
needed to reach a height of 1\,kpc during the MS lifetime of the star.
}
\label{fig:vel}
\end{figure*}

The result of the comparison of the flight times with
stellar lifetimes interpolated from the main sequence tracks shows that both
are consistent within errors for the vast majority of the sample. This is in
agreement with the previous findings and confirms that the BEM and DEM disc
ejection scenarios are capable of explaining most of the early type MS stars
found in the Galactic halo. Three notable exceptions are SB\,357,
EC\,20252$-$3137 and Hip\,77131 for which the derived flight times exceed the
evolutionary lifetimes with high statistical significance. Possible
explanations are 1) a blue straggler scenario in which a close binary is
ejected and merges afterwards or 2) the in situ formation in the Galactic
halo mentioned earlier. 

\begin{figure}[t!]
\resizebox{\hsize}{!}{\includegraphics[clip=true]{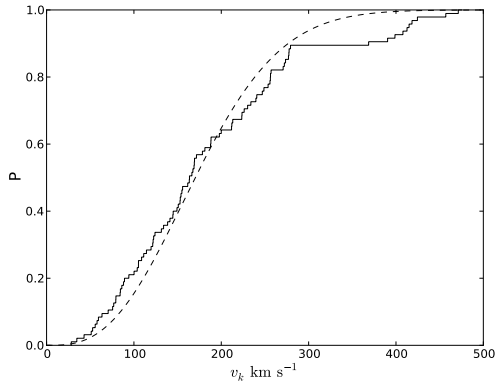}}
\caption{\footnotesize
Cumulative ejection velocity distribution.The dashed lines is the fit of a 
Maxwellian distribution function -- demonstrating the sudden break in the
distribution above 300\kms. 
}
\label{fig:cum_vel}
\end{figure}

Figure~\ref{fig:vel} shows the derived ejection velocities relative to the 
SoR at the point of origin as function of stellar mass.  
This figure shows a smooth distribution of velocities from a lower limit up to
300\,\kms. Separated by a gap a
group of ten stars with ejection velocities of $\approx$400\,\kms\ is found. 
The lower limit results from the selection of the \citet{S.N2011}
sample from early type stars above the Galactic disc.  
Could this be due to a
statistical fluke? To address this we constructed the cumulative distribution
shown in Fig.~\ref{fig:cum_vel} and compared it to the best fitting Maxwellian
distribution. While the ``low velocity'' part up to 300\,\kms\ is well
reproduced, the high velocity stars are clearly not part of a continuous
distribution, but form a separate group. We will discuss a possible link to
the hypervelocity stars in the next section.

\section{Hyper-runaway stars}

Starting in 2005 a group of stars was discovered, which have Galactic
rest-frame velocities in excess of the (local) Galactic escape velocity of
about 550\,\kms\ \citep{B.G.K2005, E.N.H2005, B.G.K2007}. Immediately after
discovery it was ``clear'' that these have been formed by the interaction of
(binary) stars with the supermassive black hole in the centre of our Galaxy
as proposed by \citet{Hil1988}.
 
However, \citet{H.E.N2008} presented the example of HD\,271791 -- a
hyper-runaway star. HD\,271791 is a B-type supergiant found far below the
Galactic plane. Its Galactic rest-frame velocity is uncertain, but a lower
limit of 530\,\kms\ has been determined, which exceeds the escape velocity at
its location, fulfilling the definition of a hypervelocity star. HD\,271791
could be traced back to its birthplace
in the Galactic plane using the methods discussed above. Uncertainties are
large, but a Galactic centre origin can be ruled out at a high significance
level. This hypervelocity star originated in the outer regions of the Galactic 
disc. A more appropriate name
for this class of star would thus be hyper-runaway stars. The ejection velocity
of HD\,271791 was $\approx$400\,\kms, placing it within the group of
high-velocity runaway stars in Fig.~\ref{fig:vel}.

\citet{P.N.H2008} carried out an abundance analysis of HD\,271791. The
observed abundance patterns are best explained by contamination of the star by
ejecta from a SN or hypernova (HN). \citet{P.N.H2008} propose a scenario in
which HD\,271791 was born in a binary with a high mass companion ($\ge
55M_\odot$). When the binary evolves and expands a common envelope is formed
causing a further contraction of the orbit. After the ejection of the common
envelope 
the high mass component has become a Wolf-Rayet (WR) star. Since
WR stars are more compact than MS stars the binary orbit can be closer
(orbital velocity $\approx$400\,\kms) than possible in an MS+MS binary. When
eventually the massive star explodes as a SN or HN the remaining
MS star is ejected with approximately its orbital velocity. Depending on the
alignment with the orbit of the binary around the Galactic centre both
velocities can add up to Galactic rest frame velocities of up to 650\,\kms,
exceeding the escape velocity.

\section{Galactic populations of runaway stars and HMXBs}

Defining a sample of runaway stars with well known selection criteria is
difficult. \citet{S.N2011} constructed a reasonably complete sample of known
high Galactic latitude runaway stars within a cylinder of radius 1\,kpc
around the Sun. Low velocity runaway stars from \citet{M.C2005} were added to
overcome the bias visible in Fig.~\ref{fig:vel}. The result of this exercise
was an estimated surface number density at the position of the Sun of
13.4\,kpc$^{-2}$.

Since we want to compare the populations of runaway stars and HMXBs, we are
interested in an estimate of the total Galactic population. A {\it crude}
estimate can be derived by scaling the surface density of the solar
neighbourhood to the surface area of the Galactic disc. Adopting a truncation
radius of 15\,kpc \citep{R.R.E1996} we arrive at a total population of 9500
runaway stars. 
 \citet{D.D.B2005}
estimated that the sample of runaway stars is consistent with a contribution
of up to 70\% produced via the supernova mechanism (BEM). Thus a back of the
envelope estimate of the Galactic runaway stars produced via the BEM channel
is 4000\ldots 5000. A further reduction to about 1500 results
from the fact that many HMXBs contain O and early B components.

This compares to a Galactic sample of known HMXBs of 114
\citep{L.vP.vdH2006}. HMXBs can -- in principle -- be detected in hard X-rays
throughout the whole Galaxy. However, some HMXBs remain in low states, in
which they are likely to escape detection by surveys, for a long time. This
leaves quite some room for systems still waiting for detection. Nevertheless,
even taking into account the large uncertainties of these estimates, it
appears that the number of HMXBs is small compared to the number of runaway
stars produced by the BEM channel. This is consistent with the expectation
that only a small minority of binary systems survived the explosion of the
more massive component.

For a significant improvement of these estimates detailed modelling will be
needed taking into account
the variation of the star formation rate in the Galaxy, travel of runaway
stars and HMXBs to other parts of the Galaxy and a number of selection effects.

\section{Conclusions}

A systematic investigation of the sample of known runaway stars shows that
the vast majority of them can be explained by disc ejection. The velocity
distribution shows a bimodality with a continuous distribution of stars up to
ejection velocities of 300\,\kms. These are explained by the standard BEM and
DEM scenarios. A group of 10 stars with ejection velocities of
$\approx$400\,\kms\ is probably explained by a variant of the BEM in which the
massive companion explodes as WR star after a common envelope phase. This
mechanism is able to produce hyper-runaway stars exceeding the Galactic
escape velocity. Estimates of the Galactic populations of runaway stars and
HMXBs are consistent with the picture that HMXBs are the small minority of the
binaries which survives the supernova explosion of the more massive companion.

 


\bigskip
\bigskip
\noindent {\bf DISCUSSION}

\bigskip
\noindent {\bf DANIELE FARGION:} To give a kick to a runaway star (among other
effects) are the neutrino explosion (10\% of NS mass) a reason for the escape?

\bigskip
\noindent {\bf RN:} Yes, mostly indirectly through their
contribution to the envelope ejection and the NS kick. 

\bigskip
\noindent {\bf VALENTI BOSCH-RAMON:} How many massive (runaway) star bow shocks 
are in the Galactic plane?

\bigskip
\noindent {\bf RN:} Most high velocity runaway stars will leave the gas layer
of the disc very fast, unless ejected almost parallel to the plane. An order
of magnitude estimate can be derived from the 5\,kpc$^{-2}$ (mostly) low velocity 
runaway stars of \citet{M.C2005}.

\end{document}